\documentclass[twocolumn]{jpsj2} 

\title{$^{75}$As NQR/NMR Studies on Oxygen-deficient Iron-based  Oxypnictide Superconductors LaFeAsO$_{1-y}$ ($y=0,0.25,0.4$) and NdFeAsO$_{0.6}$}

\author{Hidekazu Mukuda$^{1}$\thanks{E-mail address: mukuda@mp.es.osaka-u.ac.jp}, Nobuyuki Terasaki$^{1}$, Hiroaki Kinouchi$^{1}$, Mitsuharu Yashima$^{1}$, Yoshio Kitaoka$^{1}$,\\ Shinnosuke Suzuki$^{2}$, Shigeki Miyasaka$^{2}$, Setsuko Tajima$^{2}$, Kiichi Miyazawa$^{3}$, Parasharam Shirage$^{3}$, \\ Hijiri Kito$^{3}$, Hiroshi Eisaki$^{3}$, Akira Iyo$^{3}$ }

\inst{$^{1}$Graduate School of Engineering Science, Osaka University, Toyonaka, Osaka 560-8531 \\
$^2$Department of Physics, Graduate School of Science, Osaka University, Osaka 560-0043\\
$^3$National Institute of Advanced Industrial Science and Technology (AIST), Umezono, Tsukuba 305-8568 \\
}

\abst{
We report $^{75}$As-NQR/NMR studies on the oxygen-deficient iron(Fe)-based oxypnictide superconductors LaFeAsO$_{0.6}$ ($T_c=$28 K) along with the results on LaFeAsO, LaFeAsO$_{0.75}$($T_c=$20 K) and NdFeAsO$_{0.6}$($T_c=$53 K). 
Nuclear spin-lattice relaxation rate $1/T_1$ of $^{75}$As NQR at zero field on LaFeAsO$_{0.6}$ has revealed a $T^3$ dependence below $T_c$ upon cooling without the coherence peak just below $T_c$, evidencing the unconventional superconducting state with the line-node gap. We have found an intimate relationship between the nuclear quadrupole frequency $\nu_Q$ of $^{75}$As and $T_c$ for four samples used in this study. 
It implies microscopically that the local configuration of Fe and As atoms is significantly related to the $T_c$ of the Fe-oxypnictide superconductors, namely, the $T_c$ can be enhanced up to 50 K when the local configuration of Fe and As atoms is optimal, in which the band structure may be also optimized through the variation of hybridization between As $4p$ orbitals and Fe $3d$ orbitals.  
} 

\kword{superconductivity, iron-based oxypnictide, LaFeAsO, NdFeAsO, NQR, NMR}

\begin{document}

\maketitle

\date{\today}

Recent discovery of superconductivity (SC) in iron(Fe)-based transition metal oxypnictides has provided a new material base for searching high temperature superconductivity.
The structure of mother materials contains alternate stacking of RE$_2$O$_2$ and Fe$_2$Pn$_2$ layers (RE: Rare earth, Pn: Pnictogen) along the c-axis where Fe ion occupies the center of the Pn tetrahedron. 
LaFePO exhibits a superconducting transition at $T_c\sim$ 4 K and a partial replacement of O$^{2-}$ by F$^-$ increases $T_c$ up to 7 K. \cite{Kamihara2006}  The succeeding discovery of superconductivity in LaFeAsO$_{1-x}$F$_x$ reaching $T_c$=26 K \cite{Kamihara2008} has attracted much interest in the research field of superconductivity.  Furthermore, it was reported that $T_c$ increases up to 43 K by applying pressure ($P$) on LaFeAsO$_{0.89}$F$_{0.11}$ \cite{Takahashi} and exceeds 50 K when La was totally replaced by Sm even at ambient pressure ($P=0$).\cite{SmOFFeAs} The superconductivity also takes place in fluorine-free oxygen-deficient compounds REFeAsO$_{1-y}$; for example, NdFeAsO$_{0.6}$ at $T_c=54$ K\cite{Iyo} and GdFeAsO$_{0.6}$ at $T_c=53.5$ K.\cite{GdFeAsO}  In particular, a very sharp superconducting transition in resistance with increasing pressure for NdFeAsO$_{0.6}$ ensures the homogeneous electronic state even in the oxygen-deficient sample.\cite{Takeshita}  
In this letter, we report on the superconducting and normal-state properties of LaFeAsO$_{0.6}$ with $T_c=$28 K by means of $^{75}$As-NQR/NMR. We present firm evidence of the unconventional nature of the superconductivity with line-node gap for LaFeAsO$_{0.6}$ with $T_c=$28 K and an intimate relationship between a nuclear quadrupole frequency $\nu_Q$ of $^{75}$As-NQR and $T_c$ for LaFeAsO, LaFeAsO$_{0.75}$ ($T_c=$20 K), LaFeAsO$_{0.6}$($T_c=$28 K), and NdFeAsO$_{0.6}$ ($T_c=$53 K). 

Polycrystalline sample of LaFeAsO was made in the silica tubes at ambient pressure $P=0$ using LaAs, Fe and Fe$_2$O$_3$ as starting materials. An anomaly at $\sim$150 K was confirmed in the resistivity measurement, consistent with the first report by Kamihara {\it et al}.\cite{Kamihara2008} 
The polycrystalline samples of the fluorine-free oxygen-deficient compounds; LaFeAsO$_{0.75}$, LaFeAsO$_{0.6}$ and NdFeAsO$_{0.6}$ were synthesized via the high-pressure synthesis technique as described elsewhere.\cite{Iyo} 
The oxygen contents of the samples are nominal (intended) values. 
Although the real oxygen content of the samples may be larger than the nominal values due to the oxidation of the starting rare earth elements, powder X-ray diffraction measurements indicate that these samples are composed of almost a single phase. 
The lattice parameters are $a=$4.029\AA\  and $c=$8.7143\AA\  for LaFeAsO$_{0.75}$, $a=$4.022\AA\  and $c=$8.711\AA\  for LaFeAsO$_{0.6}$, which are shorter than those for the nondoped LaFeAsO\cite{Kamihara2008}. 
The lattice parameters for NdFeAsO$_{0.6}$ are $a=$3.946\AA\  and $c=$8.536\AA. 
The $T_c$ was determined by the susceptibility measurement, which exhibits a marked decrease due to an onset of superconductivity below $T_c=$ 20 K, 28 K and 53 K for LaFeAsO$_{0.75}$, LaFeAsO$_{0.6}$ and NdFeAsO$_{0.6}$, respectively. The samples were moderately crushed into powder for the NQR/NMR measurements. The $^{75}$As-NQR/NMR measurements have been performed by using the phase coherent pulsed NMR/NQR spectrometer in the temperature ($T$) range between 4 K and 280 K. The $1/T_1$ was measured with the saturation recovery method.

\begin{figure}[htbp]
\begin{center}
\includegraphics[width=0.9\linewidth]{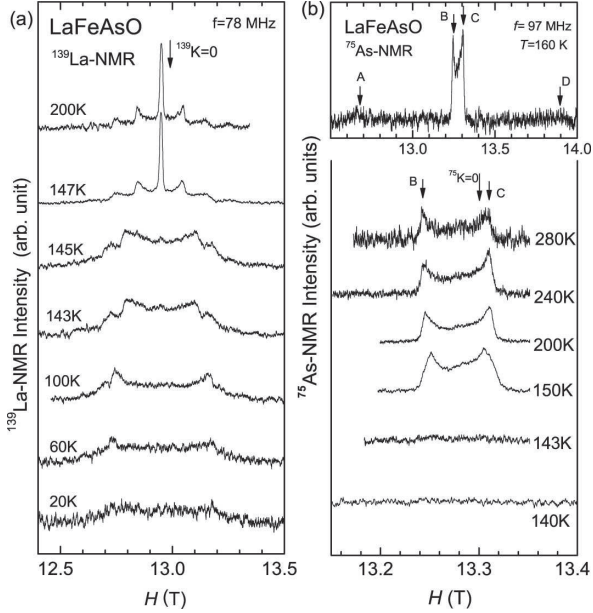}
\end{center}
\caption[]{(a) $^{139}$La-NMR spectra and (b) $^{75}$As-NMR spectra for the powder sample of LaFeAsO. The well articulated spectra in temperatures  higher than $T_N\sim$ 145 K becomes broader upon cooling below $T_N$, exhibiting a rectangular shape due to the uniform internal field induced by the AFM moments at Fe site. The $^{75}$As-NMR spectra completely disappears because of the larger internal field than at the La site, suggesting a strong mixing effect between As-$4p$ orbitals and Fe-$3d$ orbitals. These results provide firm evidence of the commensurate AFM order in LaFeAsO.}
\label{fig:LaFeAsO}
\end{figure}

Figure \ref{fig:LaFeAsO} shows the $^{75}$As-NMR (nuclear spin $I=$3/2) and $^{139}$La-NMR($I=$7/2) spectra of LaFeAsO under a magnetic field $H\sim$ 13 T.  It is evident that the well articulated $^{139}$La-NMR spectra due to the nuclear quadrupole interaction (NQI) suddenly becomes broader just below 145 K and changes into a rectangular-like spectral shape at low temperatures. This characteristic spectrum points to the appearance of uniform internal field $H_{\rm int}$ at the La site in association with a commensurate antiferromagnetic (AFM) order. The $H_{\rm int}$ at La site, that is induced by AFM moments at Fe site, is estimated to be about 0.25 T at 20 K. These results are consistent with the previous $^{139}$La-NMR studies by Nakai {\it et al}.\cite{Ishida}  Likewise, the well articulated $^{75}$As-NMR spectrum was observed above 150 K, as shown in Fig. \ref{fig:LaFeAsO}(b).  However, the $^{75}$As-NMR spectrum disappears suddenly below 145 K, which suggests that the $H_{\rm int}$ at the As site is larger than at the La site as a result of a strong mixing effect between As-$4p$ orbitals and Fe-$3d$ orbitals. 

\begin{figure}[htbp]
\begin{center}
\includegraphics[width=0.95\linewidth]{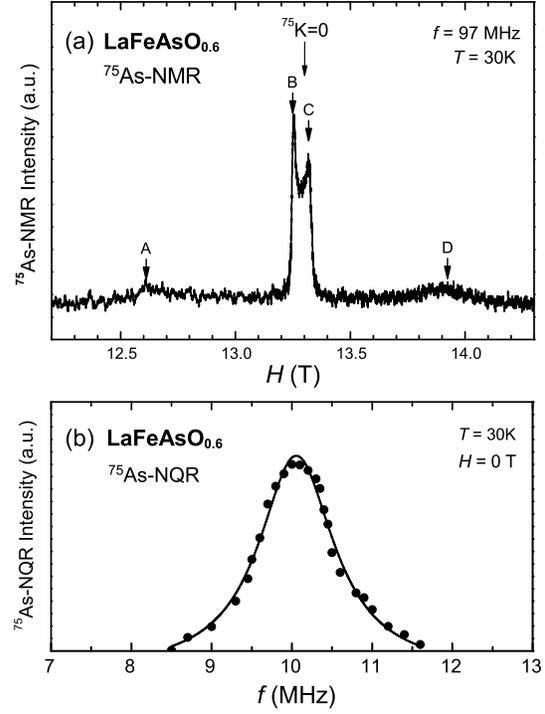}
\end{center}
\caption[]{(a) $^{75}$As NMR spectra at 30 K for the powder sample of LaFeAsO$_{0.6}$ with $T_c=28$ K. The spectra are well articulated over the whole temperature range due to the nuclear quadrupolar interaction, exhibiting the lack of AFM order.  (b)The $^{75}$As-NQR spectrum with a peak frequency at 10.05 MHz and a spectral linewidth of about 1 MHz.}
\label{fig:spectra}
\end{figure}

Figure \ref{fig:spectra}(a) shows an $^{75}$As-NMR spectrum for LaFeAsO$_{0.6}$ obtained by sweeping the magnetic field ($H$) at a fixed frequency 97 MHz and $T=30$ K. The spectrum is well articulated in 4 K$\le T \le$ 250 K due to the NQI. The central transition ($+1/2\leftrightarrow -1/2$) splits into two peaks by the second-order NQI; The left and right edges denoted as B and C in Fig. \ref{fig:spectra}(a) correspond to the resonance fields for the powders where the respective $\theta$s that is an angle between the c-axis and the external field are $\theta=90^\circ$ and 41.8$^\circ$. 
Two satellite peaks ($\pm1/2 \leftrightarrow \pm3/2$) denoted as A and D in Fig. \ref{fig:spectra}(a) arise from the first-order NQI.  
From the analysis of NQI in the spectrum, the $^{75}$As NQR frequency is estimated to be about 10 MHz. Actually, as shown in Fig. \ref{fig:spectra}(b), the $^{75}$As-NQR spectrum was observed around 10.05 MHz with a spectral width of $\sim$1 MHz.

Nuclear spin-lattice relaxation rate $(1/T_1)$ of $^{75}$As-NQR on LaFeAsO$_{0.6}$ was measured at $f=$10.05 MHz and zero field. A recovery curve of $^{75}$As nuclear magnetization ($I=3/2$) is expressed by a simple exponential function given by, 
\[
m(t)\equiv\frac{M(\infty)-M(t)}{M(\infty)}=\exp\left(-\frac{3t}{T_{1}}\right),
\]
where $M(\infty)$ and $M(t)$ are the respective nuclear magnetizations for the thermal equilibrium condition and at a time $t$ after the saturation pulse. If a system were homogeneous,  $m(t)$ should follow the simple exponential curve. Note that the $m(t)$ of $^{75}$As-NQR in LaFeAsO$_{0.6}$  was almost fitted by a single exponential function in the superconducting state and the normal state. 

\begin{figure}[htbp]
\begin{center}
\includegraphics[width=0.95\linewidth]{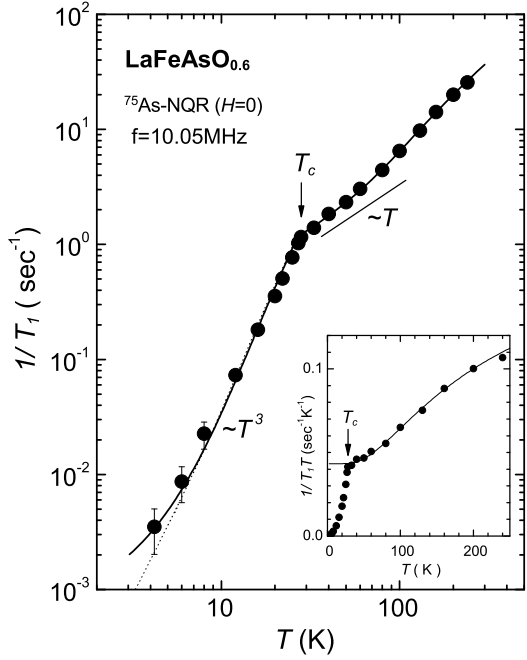}
\end{center}
\caption[]{Temperature dependence of $^{75}$As-NQR $1/T_{1}$ for LaFeAsO$_{0.6}$ at zero field. In the normal state, as shown in the inset, the $1/T_{1}T$ decreases upon cooling down to $T_c=28$ K, suggesting the significant reduction of the DOS at the Fermi level. The solid curve in the normal state is a tentative fitting curve of $1/T_1T=0.043+0.15\exp(-196/T)$, which assume a pseudo gap of $\Delta_{\rm PG}/k_B\approx 196$ K. In the superconducting state, the $1/T_{1}$ follows a $T^3$ dependence upon cooling without the coherence peak just below $T_{\rm c}=$ 28 K, evidencing the unconventional superconductivity with the line-node gap. The solid (dashed) line in the superconducting state is the calculation by assuming the line-node model where $\Delta=\Delta_0\cos{2\phi}$ with $2\Delta_0/k_BT_c\approx$5 and RDOS of $\sim$10\% (0\%).
}
\label{fig:T1}
\end{figure}

Figure \ref{fig:T1} shows the temperature dependence of $^{75}$As-NQR $1/T_{1}$ for LaFeAsO$_{0.6}$ with $T_c$=28 K. 
In the normal state, as displayed in the inset, $1/T_1T$ gradually decreases upon cooling down to $T_c$=28 K. This reduction in $1/T_1T$ probes a decrease of the density of states (DOS) at the Fermi level, resembling the pseudo-gap behavior as reported for the F-substituted compounds LaFeAsO$_{1-x}$F$_{x}$ ($x=$0.08-0.11) in the literatures.\cite{Ishida,Imai,Curro} 
If we assume such  $T$ variation as $1/T_1T= a+b\exp(-\Delta_{\rm PG}/k_BT)$, the pseudo-gap temperature $\Delta_{\rm PG}/k_B\approx 196$ K was tentatively evaluated with $a\approx0.043$ and $b\approx0.15$. The $\Delta_{\rm PG}$ is slightly larger than in the F-substituted LaFeAsO$_{0.89}$F$_{0.11}$.\cite{Ishida} 
In the superconducting state, $1/T_{1}$ shows a $T^3$ dependence without a coherence peak just below $T_{\rm c}=$28 K, giving evidence for an unconventional superconducting nature with the line-node gap. 
In the unconventional superconducting state with the line-node gap, we used to observe a $T_1T=const.$ behavior well below $T_{\rm c}$ as a result of a deviation from $T^3$ dependence. This is because the residual density of states (RDOS) at the Fermi level is induced by such an impurity effect as lattice defects due to some off-stoichiometry.
Note that such a $T_1T=const.$-like behavior was not seen down to 4 K, suggesting that the RDOS is anticipated to be less than 10\% of $N_0$. 
By assuming a line-node gap model with $\Delta=\Delta_0\cos{2\phi}$, $1/T_{1}T$ in the superconducting state is well reproduced by
\[
\frac{1/T_{1}T}{1/T_{1}T_{\rm c}}=\frac{2}{k_{\rm B}T}\int\left(\frac{N_{\rm s}(E)}{N_0}\right)^2 f(E)[1-f(E)]dE,
\]
where $N_{\rm s}(E)/N_0=E/\sqrt{E^2-\Delta^2}$; $N_0$ is the DOS at $E_{\rm F}$ in the normal state and $f(E)$ is the Fermi distribution function. 
As drawn by solid and dashed lines in Fig. \ref{fig:T1}, the $1/T_1$ data seem to be reproduced by assuming $2\Delta_0\approx 5k_{\rm B}T_{\rm c}$ and  either RDOS of 10\% or 0\%, respectively. 
The present results on the oxygen-deficient sample resemble the $^{75}$As-NMR $T_1$ results on the F-substituted LaFeAs(O$_{1-x}$F$_{x}$) with $x=0.11$ by Nakai {\it et al.}\cite{Ishida} and with $x=0.1$ by Grafe {\it et al.},\cite{Curro} although those experiments were performed in the superconducting mixed state under $H$ and hence $H$ may induce the RDOS due to the Doppler effect within the vortex core (Volovic effect)\cite{Volovik}. 
A different behavior in the $T$ dependence of $1/T_1$ was reported on PrFeAsO$_{0.89}$F$_{0.11}$ by $^{19}$F-NMR-$1/T_1$, suggesting the presence of two superconducting gaps in the framework of the line-node gap model.\cite{Matano} 
In contrast, as shown in Fig. \ref{fig:T1}, $^{75}$As-NQR $1/T_{1}$ data follow a simple $T^3$ dependence upon cooling without the coherence peak just below $T_c$, which differs from the results under $H$ reported in the literatures.\cite{Ishida,Curro,Matano} We remark that when $T_1$ is measured under $H$, the vortex core and the anisotropy of $H_{c2}$ in the superconducting mixed state may force to complicate a relaxation process to identify a superconducting characteristics. In this context, the $^{75}$As-NQR $1/T_1$ at $H$=0, which follows a simple $T^3$ dependence, is expected to unravel the superconducting characteristics inherent to the Fe-oxypnictide superconductor. 

\begin{figure}[htbp]
\begin{center}
\includegraphics[width=0.95\linewidth]{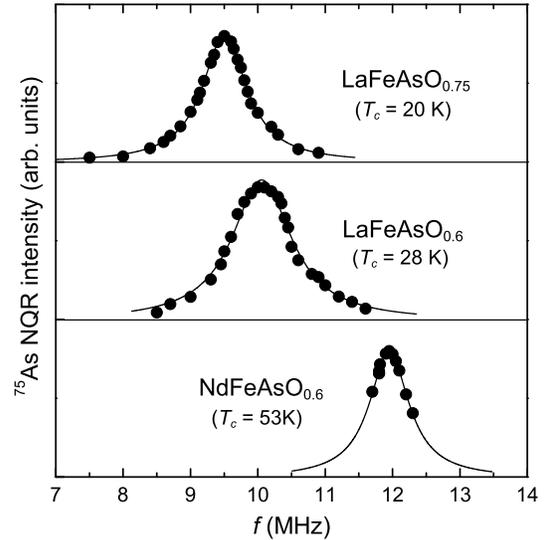}
\end{center}
\caption[]{$^{75}$As NQR spectra for LaFeAsO$_{0.75}$ ($T_c=$ 20 K), LaFeAsO$_{0.6}$ ($T_c=$ 28 K) and NdFeAsO$_{0.6}$ ($T_c=$ 53 K) observed at low temperatures. Remarkably, it was found that the $T_c$ becomes higher in the samples with larger $^{75}\nu_Q$ at As site.}
\label{fig:AsNQR}
\end{figure}

\begin{figure}[htbp]
\begin{center}
\includegraphics[width=1\linewidth]{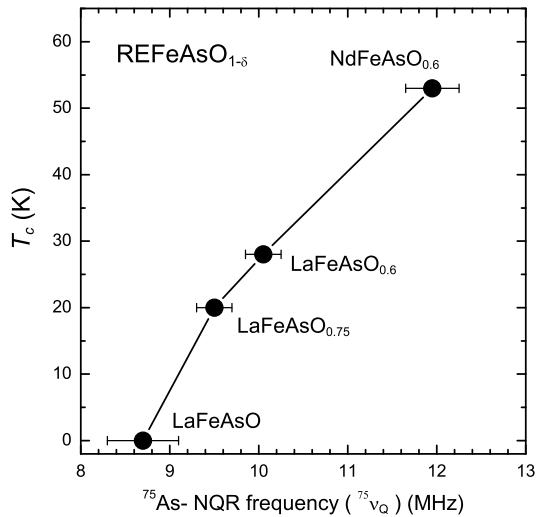}
\end{center}
\caption[]{A plot of $^{75}$As NQR frequency ($^{75}\nu_Q$) versus $T_c$ for LaFeAsO, LaFeAsO$_{0.75}$, LaFeAsO$_{0.6}$, and NdFeAsO$_{0.6}$. We have found an intimate relationship between the nuclear quadrupole frequency $\nu_Q$ of $^{75}$As and $T_c$ for four samples used in this study. 
It implies microscopically that the enhancement of $T_c$ in the Fe-oxypnictide superconductors is related to the optimal local configuration of Fe and As atoms, which may bring about the optimal band structure derived from the hybridization between As $4p$ orbitals and Fe $3d$ orbitals as well. 
}
\label{fig:NQR-Tc}
\end{figure}

Finally, we compare $^{75}$As-NQR frequencies ($^{75}\nu_Q$) of the samples used in this study. 
As shown in Fig. \ref{fig:AsNQR}, we determined the $^{75}\nu_Q$ values from the $^{75}$As NQR spectra as 9.5($\pm0.2$)MHz, 10.05($\pm0.2$)MHz and 11.95($\pm0.3$) MHz for LaFeAsO$_{0.75}$, LaFeAsO$_{0.6}$ and NdFeAsO$_{0.6}$, respectively. 
We present in Fig. \ref{fig:NQR-Tc} an experimental relationship between $T_c$ and $^{75}\nu_Q$ for LaFeAsO, LaFeAsO$_{0.75}$, LaFeAsO$_{0.6}$ and NdFeAsO$_{0.6}$. 
Here the $\nu_Q\sim 8.7(\pm0.4)$ MHz for nondoped LaFeAsO was extracted from the As-NMR spectrum well articulated in association with the NQI.
Remarkably, it was found that the $T_c$ becomes higher in the samples with larger $^{75}\nu_Q$ at As site.
It should be noted that $^{75}\nu_Q$ is proportional to the electric field gradient (EFG) along the c-axis $V_{zz}$, where $\nu_Q=eQV_{zz}/2\sqrt{1+\eta^2/3}$, the nuclear quadrupole moment $Q$ of $^{75}$As and the asymmetry parameter $\eta$ of EFG. 
The EFG is generally given by two contributions; one is non cubic charge distribution of $4p$ orbitals of  the $^{75}$As atom and the other is the charge distribution arising from the ions surrounding As, denoted as $V_{zz}^{\rm in}$ and $V_{zz}^{\rm out}$, respectively. 
The former $V_{zz}^{\rm in}$ originates from the hybridization between the As-$4p$ orbitals and Fe-$3d$ orbitals in the FeAs layer, and the latter $V_{zz}^{\rm out}$ may have a predominant contribution relevant with the Madelung potential from the charge distributions of the neighboring Fe atoms and REO$_{1-y}$ layers. 
The variation of lattice parameters significantly influences the $V_{zz}^{\rm out}$. 
Actually, the lengths of a-axis and c-axis in the tetragonal structure decrease with decreasing the oxygen contents in LaFeAsO$_{1-y}$, and further decrease by replacing La with Nd in REFeAsO$_{0.6}$. 
Despite of the reduction of the lattice volume, the recent neutron diffraction experiment by Lee {\it et al.} has revealed that a distance between the Fe- and As-plane becomes larger in NdFeAsO$_{0.6}$ than in LaFeAsO$_{0.6}$ \cite{Lee}. 
The simple calculation of $V_{zz}^{\rm out}$ by assuming the point charges at the atoms surrounding As revealed that the $V_{zz}^{\rm out}$ is larger for NdFeAsO$_{0.6}$ than for LaFeAsO$_{0.6}$ and LaFeAsO. 
However, it was not proportional to the $^{75}\nu_Q$ precisely, indicating that the on-site contribution  $V_{zz}^{\rm in}$ is also important in these compounds, namely, the charge distribution at As-$4p$ orbitals also varies from non-superconducting LaFeAsO to NdFeAsO$_{0.6}$ with $T_c=$53 K.  
It may be derived from the variation of the hybridization between As-$4p$ orbitals and Fe-$3d$ orbitals, which brings about the modification of the Fe-As layer-derived band structure as well. 
As a result, the intimate relation between $\nu_Q$ and  $T_c$ suggests that the local configuration of Fe and As atoms is significantly related to the $T_c$ of the Fe-oxypnictide superconductors, in other words, the $T_c$ can be enhanced up to 50 K when the local configuration of Fe and As atoms is optimal, in which the band structure may be optimized for the superconducting state through the variation of hybridization between As $4p$ orbitals and Fe $3d$ orbitals as well. 
Interestingly, Lee {\it et al.} pointed out that the $T_c$ becomes maximum when the bonding angle between the Fe-As-Fe coincides with that of a regular tetrahedron of FeAs$_4$\cite{Lee}.
We remark that the further analysis on $\nu_Q$ is important to reveal the origin of superconductivity in the Fe-oxypnictide compounds.

In summary, we revealed an onset of antiferromagnetic order below 145 K in nondoped LaFeAsO by means of $^{139}$La- and $^{75}$As-NMR, confirming the previous results. 
The $^{75}$As-NQR $1/T_1$ measurements have established that the LaFeAsO$_{0.6}$ with $T_c=$ 28 K is the unconventional superconductor with the line-node gap with $2\Delta/k_BT_c\approx$5.  The systematic measurement of $^{75}$As nuclear quadrupole frequency ($^{75}\nu_Q$) on the nondoped LaFeAsO($T_c=0$ K), LaFeAsO$_{0.75}$($T_c=20$ K), LaFeAsO$_{0.6}$($T_c=28$ K) and  NdFeAsO$_{0.6}$($T_c=53$ K) has revealed the intimate relationship between $^{75}\nu_Q$ and $T_c$. 
This result implies microscopically that the enhancement of $T_c$ in the Fe-oxypnictide superconductors is strongly related to the optimal local configuration of Fe and As atoms, and the optimal band structure derived from the hybridization between As $4p$ orbitals and Fe $3d$ orbitals. 




\begin{thebibliography}{99} 

\bibitem{Kamihara2006} Y. Kamihara, H. Hiramatsu, M. Hirano, R. Kawamura, H. Yanagi, T. Kamiya and H. Hosono: J. Am. Chem. Soc. {\bf 128} (2006) 10012.
\bibitem{Kamihara2008}  Y. Kamihara, T. Watanabe, M. Hirano and H. Hosono: J. Am. Chem. Soc. {\bf 160} (2008) 3296.
\bibitem{Takahashi} H. Takahashi, K. Igawa, K. Arii, Y. Kamihara, M. Hirano and H. Hosono: Nature {\bf 453} (2008) 376.
\bibitem{SmOFFeAs} Z. A. Ren, W. Lu, J. Yang, W. Yi, X. L. Shen, Z. C. Li, G. C. Che, X. L. Dong, L. L. Sun and  F. Zhou: Chin. Phys. Lett. {\bf 25} (2008) 2215.
\bibitem{Iyo} H. Kito, H. Eisaki and A. Iyo: J. Phys. Soc. Jpn. {\bf 77} (2008) 063707.
\bibitem{GdFeAsO} J. Yang, Z.C. Li, W. Lu, W. Yi, X.L. Shen, Z.A. Ren, G.C. Che, X.L. Dong, L.L. Sun, F. Zhou and Z.X. Zhao: Supercond. Sci. Technol. {\bf 21} (2008) 082001.
\bibitem{Takeshita} N. Takeshita, A. Iyo, H. Eisaki, H. Kito and T. Ito: J. Phys. Soc. Jpn. {\bf 77} (2008) 075003. 
\bibitem{Ishida} Y. Nakai, K. Ishida, Y. Kamihara, M. Hirano and H. Hosono: J. Phys. Soc. Jpn. {\bf 77} (2008) 073701.
\bibitem{Imai} K. Ahilan, F. L. Ning, T. Imai, A. S. Sefat, R. Jin, M. A. McGuire, B. C. Sales and D. Mandrus: arXiv:0804.4026.
\bibitem{Curro}	H.-J. Grafe, D. Paar, G. Lang, N. J. Curro, G. Behr, J. Werner, J. Hamann-Borrero, C. Hess, N. Leps, R. Klingeler and  B. Buechner: arXiv:0805.2595. 
\bibitem{Matano} K. Matano, Z. A. Ren, X. L. Dong, L. L. Sun, Z. X. Zhao, and G.-q. Zheng: to be published in J. Phys. Soc. Jpn. {\bf 77} (2008).
\bibitem{Volovik} G. E. Volovik: JETP Lett. {\bf 58} (1993) 469.
\bibitem{Lee} C. H. Lee, T. Ito, A. Iyo, H. Eisaki, H. Kito, M. T. Fernandez-diaz, K. Kihou, H. Matsuhata, M. Braden and K. Yamada: to be published in J. Phys. Soc. Jpn.

\end{thebibliography}
\end{document}